\documentstyle[preprint,aps]{revtex}
\title{A Non-Singular  Universe in String  Cosmology}
\author{ A. A. Al-Nowaiser, Murat \"Ozer and  M. O. Taha}
\address{ Department of Physics, College of Science,
 King Saud University, P. O. Box 2455, Riyadh 11451, Saudi Arabia}
\begin{document}
\draft
\maketitle

\begin{abstract}
We consider the low-energy effective string action in four dimensions including the leading order-$\alpha'$  terms. An exact homogeneous solution is obtained. It represents a non-singular expanding cosmological model in which the  tensor fields tend to vanish as $t \rightarrow \infty$. The scale factor $a(t)$ of the very early universe in this model has the time dependence $a(t)^2=a_0^2+t^2$. The violation of the strong energy condition of classical General Relativity to avoid the initial singularity requires that the central charge deficit of the theory be larger than a certain value. The significance of this solution is discussed.
\end{abstract}
\vspace{1.5cm}

The search for cosmological models that do not suffer from the problems inflicting the standard big bang model of cosmology has intensified since the advent of string theory. String theory is particularly relevant to the  initial singularity problem whose solution has long been thought to  require a quantum theory of gravity, for which string theory seems to be the most promising candidate. The effective action  for string theory has been calculated in the $\alpha'$-expansion  in both the sigma model approach [1-4] in which one looks upon string theory as a two dimensional conformal field theory in background fields and  the S-matrix approach [5]. Recently, modifications introduced by string theory, near the Planck  scale,  to  classical general relativity and cosmology have been considered by many physicists [6-29]. Brustein and Veneziano [22] and Kaloper, Madden and Olive [27] showed that a nonsingular inflationary universe in string theory via branch changing from a previously superexponentionally expanding phase to  a  Friedmann-Robertson-Walker  phase  is not  realized. Noninflationary singularity-free superstring cosmological solutions were found by Antoniadis, Rizos and Tamvakis [24] for a spatially flat background. Their work was extended to nonzero spatial curvature by Easther and Maeda [29]. Last, we mention the work of Gasperini, Maharana and Veneziano [15] which is rather different from the previous ones. They use the boosts of the $O(d, d)$ group to generate out of the Milne metric a one-parameter family of non-singular string cosmology solutions.

In this letter, we undertake the study of whether or not a nonsingular universe emerges in heterotic	 string theory from a point of view which is different from those of others in many ways. First of all, rather than doing numerical calculations we believe that analytic solutions to approximate equations that arise in this problem exist and are illuminating. We consider first order corrections in $\alpha'$, the inverse string tension, to the equations of motion in string theory and ask whether an exact solution with some plausible features may arise from the dynamics of this \mbox{low-energy} approximation. We find one such solution for the case of positive curvature. The scale factor of the universe in this solution has the  simple and interesting  \mbox{time-dependence} 
\begin{equation} \label{eq.1}
a(t)^{2}=a_{0}^2+t^{2}
\end{equation}
which was previously found as an exact solution to the \mbox{variable-$\Lambda$} model [29]. Here $a_{0}$ is the scale factor of the universe at $t=0$. The scalar field $\phi$, the dilaton, in this solution is given by
\begin{equation} \label{eq.2}		              			\phi=-\ln\left(\frac{a^{2}}{a_{0}^2}\right).
\end{equation}
	          
In the present paper we include the \mbox{two-loop} terms and consider \mbox{non-vanishing} \mbox{time-dependent} $H$ and $F$  tensor fields (see eq.(4)) as well as the scalar field $\phi$. We find that  the central charge term must be included and  be larger than a certain number so as to have a non-singular solution. None of the references that have obtained explicit cosmological solutions take into account  the $F-tensor$  that arises at the \mbox{two-loop} level. Our purpose is to obtain an exact homogeneous solution for the case of positive curvature. Equations (1) and (2) are characteristic features of our solution. In the present case it is sufficient to supplement  these equations  with the requirement of homogeneity to obtain an exact solution to the field equations that determines the field strenghts $H^2$ and $trF^2$. We  find 
\begin{equation} \label{eq.3}
H^2\sim a^{-6}\hspace{2mm},\hspace{2mm}trF^2\sim a^{-4}
\end{equation}
as $a\rightarrow\infty$. The first of these results is in agreement with that obtained in ref.[24]. We also construct the effective potential $V(\phi)$  that results on imposing our exact solution on the \mbox{low-energy} string action. 

To first order in the inverse string tension $\alpha'$, in the Einstein frame the effective string action in $D$ dimensions is of the form [1,2,4]
\begin{equation} %eq.(4)
I=\int d^Dx\sqrt{g}]\left[R-\frac{4}{D-2}(\nabla \Phi)^2-\frac{1}{12}H^2e^{-\frac{8\Phi}{D-2}}-\frac{1}{4}\alpha'trF^2e^{-\frac{4\Phi}{D-2}}+ce^{\frac{4\Phi}{D-2}}\right],
\end{equation}
where \footnote{We use the conventions of Weinberg [31] for metrics, curvatures, etc.}
\begin{eqnarray} \label{eq.(5)}
(\nabla \Phi)^{2}&=&g^{\mu\nu}\nabla _{\mu}\Phi\nabla_{\nu}\Phi \nonumber \\
H^2&=&H_{\mu\nu\sigma}H^{\mu\nu\sigma}\hspace{2mm},\hspace{2mm}(F^2)_{ab}=F_{a\mu\nu}F_{b}^{\mu\nu},\nonumber \\
H_{\mu\nu\sigma}&=&3\partial_{[\mu}B_{\nu\sigma]}\hspace{2mm},\hspace{2mm}F_{a\mu\nu}=\partial_{\nu}A_{a\mu}-\partial_{\mu}A_{a\nu}+gC_{abc}A_{b\mu}A_{c\nu}.
\end{eqnarray} 
We have set $8\pi G_{D}=1$, with $G_{D}$  being the $D$-dimensional gravitational constant. The field $B_{\mu\nu}$ is the antisymmetric tensor field  and $A_{a\mu}$  is the background \mbox{space-time} gauge field. The constant $c$  is the central charge deficit of the theory, 
$c=-2(D_{eff}-D_{crit})/3\alpha'$ in the heterotic and superstring theories with $D_{eff}=\frac{3}{2}D$ and $D_{crit}=15$.  The field $\Phi$ is the fundamental scalar field of string theory, the dilaton.  In the action (4) we have neglected geometric terms that are second order in $R, R_{\mu\nu}$, or $R_{\mu\nu\sigma}$. One also notes that $\nabla_{\mu}$ in (4) is the covariant \mbox{space-time} derivative in $D$-dimensions.

The action (4) may be obtained as a result of a \mbox{"low-energy"} expansion starting from the basic \mbox{two-dimensional} string action [1]. It may , alternatively, be obtained as the action whose variation produces the perturbative field equations that result from the vanishing of the string  "beta functions" [2,4]. The two approaches appear to be equivalent.

Variations of the action (4) with respect to $\Phi, g_{\mu\nu}, B_{\mu\nu}$ and $A_{a\mu}$  yield the following field equations:
\begin{equation} %eq.6
\nabla^2\Phi+\frac{1}{12}H^{2}e^{-\frac{8\Phi}{D-2}}+\frac{\alpha'}{8}trF^{2}e^{-\frac{4\Phi}{D-2}}+\frac{c}{2}e^{\frac{4\Phi}{D-2}},
\end{equation}
\begin{equation} %eq.7
R_{\mu\nu}-\frac{1}{2}g_{\mu\nu}R=-T_{\mu\nu},
\end{equation}
\begin{eqnarray} %eq.8
T_{\mu\nu}=\frac{8}{D-2}\left[-\nabla_{\mu}\Phi\nabla_{\nu}\Phi+\frac{1}{2}
(\nabla\Phi)^{2} g_{\mu\nu}\right]-
\frac{1}{2}e^{-\frac{8\Phi}{D-2}}\left[H^{2}_{\mu\nu}-\frac{1}{6}H^{2}g_{\mu\nu}\right]\nonumber \\
-\frac{\alpha'}{2}e^{-\frac{4\Phi}{D-2}}\left[2F^{2}_{\mu\nu}-\frac{1}{2}trF^{2}g_{\mu\nu}\right]-ce^{\frac{4\Phi}{D-2}}g_{\mu\nu},
\end{eqnarray}
\begin{equation} %eq.9
F^{2}_{\mu\nu}=F_{a\sigma\mu}F^{\sigma}_{a\nu}\hspace{2mm},\hspace{2mm}
H_{\mu\nu}^{2}=H^{\alpha\beta}_{\mu}H_{\alpha\beta\nu},
\end{equation}
\begin{equation} %eq.10         
\nabla_{\lambda}\left(e^{-\frac{8\Phi}{D-2}}H^{\lambda}_{\mu\nu}\right)=0,
\end{equation}
\begin{equation} %eq.11
\nabla^{\mu}\left(F_{a\mu\nu}e^{-\frac{4\Phi}{D-2}}\right)=gC_{abc}A^{\mu}_{b}F_{c\mu\nu}e^{-\frac{4\Phi}{D-2}}.
\end{equation}

In the following we shall take  $D=4$ and attempt to obtain an exact solution with a homogeneous and isotropic  \mbox{Robertson-Walker} metric given by
\begin{equation} %eq.12
ds^{2}=-dt^{2}+a(t)^{2}\left[\frac{dr^{2}}{1-kr^{2}}+r^{2}(d\theta^{2}+sin^{2}\theta d\varphi^{2})\right],
\end{equation}
so that the non-vanishing components  of $R_{\mu\nu}$  are  
\begin{equation}
R_{ij}=-\left[\frac{\ddot a}{a}+\frac{2}{a^{2}}(\dot a^{2}+k)\right]g_{ij}
\hspace{2mm},\hspace{2mm}R_{00}=3\frac{\ddot a}{a}.
\end{equation}
The field equations (7) then become:
\begin{equation} %eq.14
3\frac{\ddot a}{a}=\left(T^{0}_{0}-\frac{1}{2}T\right),
\end{equation}
\begin{equation} %eq.15
\left[\frac{\ddot a}{a}+\frac{2}{a^{2}}(\dot a^{2}+k)\right]\delta^{i}_{j}=\left(T^{i}_{j}-\frac{1}{2}T\delta^{i}_{j}
\right),
\end{equation}		                                    
where $T=T^{\mu}_{\mu}$. From eq.(15) one notes that $T^{i}_{j}$  is proportional to $\delta^{i}_{j}$, a consequence of maximal spatial symmetry.  We shall assume that, in this homogeneous universe, all fields are functions of $t$ only. Then, except for the third and fifth  terms in eq.(8), all  terms contributing to $T^{i}_{j}$  are proportional to $\delta^{i}_{j}$. It follows that the sum of these two  terms must  also be proportional to $\delta^{i}_{j}$. Now, eq.(10) may be solved by the ansatz 
\begin{equation} %eq.16
H^{\lambda\mu\nu}=\epsilon^{\lambda\mu\nu\sigma}e^{4\Phi}\nabla_{\sigma}\rho,
\end{equation}
where $\rho=\rho(t)$  is a homogeneous scalar field. Then one finds that 
\begin{equation} %eq.17
\left.H^{2}\right.^{\mu}_{\nu}=\epsilon^{\alpha\beta\mu\sigma}\epsilon_{\alpha\beta\nu\lambda} e^{8\Phi}\nabla_{\sigma}\rho\nabla_{\lambda\rho}
\end{equation}					   			     is given by 
\begin{equation} %eq.18
\left.H^{2}\right.^{i}_{j}=h(t)\delta^{i}_{j}\hspace{2mm},\hspace{2mm}\left.H^{2}\right.^{0}_{0}=0,
\end{equation}
where 
\begin{equation} %eq.19
h(t)=-2e^{8\Phi}\dot \rho^{2}.				                \end{equation}					        
Under the fact that $T^{i}_{j}\propto \delta^{i}_{j}$, it  follows from eq.(8) that $\left.F^{2}\right.^{i}_{j}$  is of the form
\begin{equation} %eq.20
\left.F^{2}\right.^{i}_{j}=f(t)\delta^{i}_{j}.
\end{equation}
This implies that the field equations (11) must be solved subject to the restriction (20) with the definite function $f(t)$ to be determined in the following exact solution. The tensor $H^{\lambda\mu\nu}$ will, however, be completely determined by our solution through equations (16) and (19). 
Next we  obtain the expressions on the \mbox{right-hand}  side in equations (14) and (15) using equations (8),(18) and (20). Towards this we note that $\nabla_{i} \Phi=0$ and set $\left.F^{2}\right.^{0}_{0}=f_{0}$  so that 
\begin{equation} %eq.21
H^{2}=3h\hspace{2mm},\hspace{2mm}trF^{2}=3f+f_{0}.
\end{equation}
One finds:
\begin{equation} %eq.22
T^{i}_{j}=\left[2(\nabla \Phi)^{2}-\frac{1}{4}e^{-4\Phi}h-
\frac{\alpha'}{4}e^{-2\Phi}(f-f_{0})-ce^{2\Phi}\right]\delta^{i}_{j},
\end{equation}
\begin{equation} %eq.23
T^{0}_{0}=4\left[\frac{1}{2}(\nabla \Phi)^{2}+\dot \Phi^{2}\right]+\frac{1}{4}
e^{-4\Phi}h+\frac{3}{4}\alpha' e^{-2\Phi}(f-f_{0})-ce^{2\Phi},
\end{equation}
\begin{equation} %eq.24
T=4(\nabla \Phi)^{2}-\frac{1}{2}e^{-4\Phi}h-4ce^{2\Phi}.
\end{equation}      	                    
Substituting these into equations (14) and (15) one obtains:
\begin{equation} %eq.25
\frac{\ddot a}{a}=\frac{1}{3}\left[4(\dot\Phi)^{2}+ \frac{1}{2}e^{-4\Phi} h-\frac{3}{4}\alpha'e^{-2\Phi}(f-f_{0})+ce^{2\Phi}\right],
\end{equation} 
\begin{equation} %eq.26
\frac{\ddot a}{a}+\frac{2}{a^{2}}(\dot a^{2}+k)=
\left[-\frac{\alpha'}{4}
e^{-2\Phi}(f-f_{0})+ce^{2\Phi}\right],
\end{equation}
The scalar field equation (6) is now
\begin{equation} %eq.27
\nabla^{2}\Phi+\frac{1}{4}e^{-4\Phi}h+\frac{\alpha'}{8}e^{-2\Phi}(3f+f_{0})+\frac{c}{2}e^{2\Phi}=0
\end{equation}                                                            
Note that in eq.(27) $\nabla^{2}\Phi=\ddot \Phi+3(\dot a/a)\dot \Phi$. Equations (25), (26) and (27) are coupled \mbox{non-linear} second order differential equations for which we seek a particular solution which is \mbox{non-singular} at $t=0$. For the case $k=1$ we were able to obtain the following exact solution in which we have set $\phi=2\Phi$:
\begin{equation} %eq.28
a=a_{0}e^{_\phi/2}\hspace{2mm},\hspace{2mm}a^{2}=a_{0}^{2}+t^{2},
\end{equation} 
\begin{equation} %eq.29
\phi=-\ln\left(1+\frac{t^{2}}{a_{0}^{2}}\right),
\end{equation} 
\begin{equation} %eq.30
h=4\left(c-\frac{8}{a_{0}^{2}}\right)e^{3\phi}+\frac{20}{a_{0}^{2}}e^{4\phi},
\end{equation} 
\begin{equation} %eq.31
\alpha'f=-2\left(c-\frac{8}{a_{0}^{2}}\right)e^{2\phi}-
\frac{11}{a_{0}^{2}}e^{3\phi},
\end{equation} 
\begin{equation} %eq.32
\alpha' f_{0}=-2\left(3c-\frac{16}{a_{0}^{2}}\right)e^{2\phi}-
\frac{15}{a_{0}^{2}}e^{3\phi},
\end{equation} 
Now we give general remarks on the significance of this solution.

(1) Note that this solution does not possess an initial singularity. This is not in contradiction with the general singularity theorems in classical General Relativity, since the strong energy condition ($SEC$) [32,33] which is sufficient for the existence of the initial singularity
\begin{equation} %eq.33
\rho+3p\geq 0 \hspace{2mm} and \hspace{2mm} \rho+p\geq 0,
\end{equation}
where $\rho=T_{00}$ and $p=T_{ii}$ (no summation over $i=1,2,3$), is avoided by a choice of $c$ within an allowed interval. Using equations (22), (23) and (29)-(32) $\rho+3p<0$ and $\rho+p<0$ give, respectively
\begin{equation} %eq.34
c>\frac{4}{a_{0}^{2}}-\frac{3}{2a_{0}^{2}}e^{\phi}\hspace{2mm},\hspace{2mm}c>\frac{14}{3a_{0}^{2}}-\frac{5}{3a_{0}^{2}}e^{\phi}.
\end{equation}
Requiring, on the other hand, that the energy density $\rho$ be positive gives
\begin{equation}%eq.35
c<\frac{6}{a_{0}^{2}}-\frac{2}{a_{0}^{2}}e^{\phi}.
\end{equation}
Thus the $SEC$ is violated and the initial singularity at $t=0$ is avoided
provided $c$, the central charge deficit, satisfies
\begin{equation}%eq.36
\frac{3}{a_{0}^{2}}<c<\frac{4}{a_{0}^{2}}.
\end{equation}

(2) As previously mentioned, one immediately observes that, whereas $\phi\rightarrow -\infty$ as $t\rightarrow\infty$, both $H^{2}$ and $trF^{2}$	tend  to vanish in this limit. This may possibly indicate that no appreciable consequences of these tensor fields can, at present, be observed.

(3) The particular solution that we have obtained is a natural generalization of the work of ref.[16], and also that of ref.[30]. It therefore possesses all the desirable properties of the models discussed in these references and  constitutes a viable \mbox{non-singular} cosmological model. In particular one notes the \mbox{time-symmetry} in the evolution of the scale factor.

(4) The solution does not exist if either $F$ or $H$ vanishes identically. 

(5) Extending the above remark, one may further observe that the dynamics of the  evolution  of  the system is  driven  by  the existence of the fields $F$ and $H$. In fact, one may write the Lagrangian density for this model in the form
\begin{equation} %eq.37
{\cal L}=R-\frac{1}{2}(\nabla \phi)^{2}-V(\phi),
\end{equation}			    			                   
where, with $8\pi G=1$,
\begin{equation} %eq.38
V(\phi)=\frac{1}{12}H^{2}e^{-2\phi}+\frac{\alpha'}{4}trF^{2}e^{-\phi}-
ce^{\phi}			           		       
\end{equation}
is the effective potential. Using our particular  solution we find $V(\phi)$ to be given by
\begin{equation} %eq.39
V(\phi)=\alpha_{1}e^{\phi}-\alpha_{2}e^{2\phi},				\end{equation}			                   
where 
\begin{equation} %eq.40
\alpha_{1}=\frac{12}{a_{0}^{2}}-2c
\hspace{2mm},\hspace{2mm}\alpha_{2}=\frac{7}{a_{0}^{2}}.
\end{equation}			         
This potential is of exactly the same form as that of ref.[15]. One should, however, note that the values of the field $\phi$ in eq.(39) are restricted to those that are allowed by the particular solution (28)-(32), i.e. to $\phi \leq 0$, as is evident from eq.(31). One should not therefore use $V(\phi)$ as given by eq.(39) in the region $\phi>0$. In particular no global properties can be deduced from eq.(39).

(6) The remark in (5) may be further generalized by observing that homogeneous solutions to the field  equations (6)-(11), based on the string action (4), may always be reexpressed with dependence on the scalar field  $\phi$ replacing that of the time parameter $t$. Equations (7) and (8) would then yield the gravitational field equations for a scalar field $\phi$  with a definite effective potential $V(\phi)$  generated by the non-scalar fields for at least the allowed region of $\phi$. This observation simplifies the dynamical considerations in obtaining cosmologically relevant solutions, since one may impose required conditions on this generated scalar potential, in its region of validity, and then restrict the solutions of equations (10) and (11) so that these conditions are satisfied. We suggest that further work on these field equations adopts such an approach.

(7) One should, in particular, note that exponential inflation is not necessary. It may indeed be noted that exponential inflation does not appear to be natural in perturbative string theory. 

One of us (M\"O) wishes to thank the Elementary Particle Physics Theory Group at the University of Maryland, College Park for the hospitality extended to him during the Summer of 97 while he was a visitor. In particular, he thanks Prof. C. H. Woo for pointing out that the scalar field potential in eq.(39) is unbounded for $\phi>0$, and related discussions.
	
{\bf References}\\
{[1]} E. S. Fradkin and A. A. Tseytlin, Phys. Lett. {\bf B 158} (1985) 316 ;
      Nucl. Phys. {\bf B 261} (1985) 1.\\
{[2]} C. G. Callan, D. Friedan , E. J. Martinec  and M. J. Perry, Nucl. Phys. {\bf B 262} (1985) 593.\\ 
{[3]} A. Sen, Phys. Rev. {\bf D 32}, 2102 (1985); Phys. Rev. Lett {\bf 55}, 1946 (1985).\\ 
{[4]} C. G. Callan, I. R. Klebanov  and M. J. Perry, Nucl. Phys. {\bf B 278}, (1986) 78.\\ 
{[5]} D. Gross and J. H. Sloan, Nucl. Phys. {\bf B 291} (1987) 41. \\
{[6]} K. Maeda, Phys. Rev. {\bf D 35} (1987) 471.\\
{[7]} I. Antoniadis, C. Bachas, J. Ellis and D. V. Nanopoulos, Phys. Lett. {\bf B 211} (1988) 393.\\
{[8]} S. Kalara, C. Kounnas, K. A. Olive, Phys. Lett. {\bf B 215} (1988) 265.\\
{[9]} N. Sanchez and G. Veneziano, Nucl. Phys. {\bf B 333} (1990) 253.\\
{[10]} G. Veneziano, Phys. Lett. {\bf B 265} (1991) 287.\\
{[11]} I. Antoniadis, C. Bachas, J. Ellis and D. V. Nanopoulos, Phys. Lett. {\bf B 257} (1991) 278.\\
{[12]} J. Wang and S. Mohanty, Phys. Lett. {\bf B 269} (1991) 279.\\
{[13]} M. C. Bento, O. Bertolami and P. M. Sa, Phys. Lett. {\bf B 262} (1991) 11.\\
{[14]} B. A. Campbell, A. Linde and K. A. Olive, Nucl. Phys. {\bf B 355} (1991) 146.\\
{[15]} M. Gasperini, J. Maharana and G. Veneziano, Phys. lett. {\bf B 272}
(1991) 277.\\
{[16]} M. \"Ozer and M. O. Taha, Phys. Rev. {\bf D 45} (1992) 997.\\
{[17]} A. A. Tseytlin, Int. J. Mod. Phys {\bf D1} (1992) 223.\\
{[18]} A. A. Tseytlin and C. Vafa, Nucl. Phys. {\bf B 372} (1992) 443.\\
{[19]} B. A. Campbell, N. Kaloper and K. A. Olive, Phys. Lett. {\bf B 277} (1992) 265.\\
{[20]} R. Brustein and P. J. Steinhadt, Phys. Lett. {\bf B 302} (1993) 196.\\
{[21]} D. S. Goldwirth and M. J. Perry, Phys. Rev. {\bf D 49} (1994) 5019.\\
{[22]} R. Brustein and G. Veneziano, Phys. Lett. {\bf B 329} (1994) 429.\\
{[23]} E. J. Copeland, A. Lahiri and D. Wands, Phys. Rev. {\bf D 50} (1994) 4868.\\
{[24]} I. Antoniadis, J. Rizos and K. Tamvakis, Nucl. Phys. {\bf B 415} (1994) 497.\\
{[25]} E. J. Copeland, A. Lahiri and D. Wands, Phys. Rev. {\bf D 51} (1995) 1569.\\
{[26]} C. Angelantonj, L. Amendola, M. Litterio and F. Occhionero, Phys. Rev. {\bf D 51} (1995) 1607.\\
{[27]} N. Kaloper, R. Madden and K. A. Olive, Nucl. Phys {\bf B 452} (1995) 677.\\
{[28]} R. Easther, K. Maeda and D. Wands, Phys. Rev. {\bf D 53} (1996) 4247.\\
{[29]} R. Easther and K. Maeda, Phys. Rev. {\bf D 53} (1996) 4247 ;Phys.Rev.D 54 (1996) 7252.\\
{[30]}  M. \"Ozer and M. O. Taha, Phys. Lett. {\bf B 171}, (1986) 363 ; Nucl. Phys. B 287 (1987) 776.\\
{[31]} S. Weinberg, {\it Gravitation and Cosmology} (Wiley, New York, 1972).\\
{[32]} S. W. Hawking and G. F. R. Ellis, {\it The Large Scale Structure of Space-Time} (Cambridge Univesity Press, Cambridge, England, 1973)\\
{[33]} M. Visser, Phys. Rev. {\bf D 56} (1997) 7578.
\end{document}